\documentclass[lettersize,journal]{IEEEtran}

\usepackage{amssymb}
\usepackage[cmex10]{amsmath}
\usepackage{stfloats}
\usepackage{graphicx}
\usepackage{subfigure}
\usepackage{tabularx}
\usepackage{epsfig,epsf,color,balance,cite}
\usepackage{epstopdf}
\usepackage{verbatim}
\usepackage{url}
\usepackage{bm}
\allowdisplaybreaks[2]

\usepackage{algorithm}
\usepackage{algorithmic}
\begin{document}

\title{A Joint Communication and Computation Design for Distributed RISs Assisted Probabilistic Semantic Communication in IIoT}

\author{Zhouxiang Zhao,~\IEEEmembership{Graduate Student Member,~IEEE,}
        Zhaohui Yang,
        Chongwen Huang,
        Li Wei,
        Qianqian Yang,
        Caijun Zhong,
        Wei Xu,
        and Zhaoyang Zhang,~\IEEEmembership{Senior Member,~IEEE}
		
    \thanks{Z. Zhao, Z. Yang, C. Huang, Q. Yang, C. Zhong, and Z. Zhang are with College of Information Science and Electronic Engineering, Zhejiang University, Hangzhou, Zhejiang, 310027, China, and Zhejiang Provincial Key Lab of Information Processing, Communication and Networking (IPCAN), Hangzhou, Zhejiang, 310007, China. Z. Yang is also with Zhejiang Lab, Hangzhou, Zhejiang, 311100, China (e-mails: \{zhouxiangzhao, yang\_zhaohui, chongwenhuang, qianqianyang20, caijunzhong, ning\_ming\}@zju.edu.cn).}
    \thanks{L. Wei is with the Engineering Product Development (EPD) Pillar, Singapore University of Technology and Design, 487372, Singapore (e-mail: weili@mymail.sutd.edu.sg).}
    \thanks{W. Xu is with the National Mobile Communications Research Lab, and Frontiers Science Center for Mobile Information Communication and Security, Southeast University, Nanjing, 210096, China, and also with Purple Mountain Laboratories, Nanjing, 211111, China (email: wxu@seu.edu.cn).}
        }
 

\maketitle

\IEEEpeerreviewmaketitle

\begin{abstract}
In this paper, the problem of spectral-efficient communication and computation resource allocation for distributed reconfigurable intelligent surfaces (RISs) assisted probabilistic semantic communication (PSC) in industrial Internet-of-Things (IIoT) is investigated. In the considered model, multiple RISs are deployed to serve multiple users, while PSC adopts compute-then-transmit protocol to reduce the transmission data size. To support high-rate transmission, the semantic compression ratio, transmit power allocation, and distributed RISs deployment must be jointly considered. This joint communication and computation problem is formulated as an optimization problem whose goal is to maximize the sum semantic-aware transmission rate of the system under total transmit power, phase shift, RIS-user association, and semantic compression ratio constraints. To solve this problem, a many-to-many matching scheme is proposed to solve the RIS-user association subproblem, the semantic compression ratio subproblem is addressed following greedy policy, while the phase shift of RIS can be optimized using the tensor based beamforming. Numerical results verify the superiority of the proposed algorithm.
\end{abstract}

\begin{IEEEkeywords}
Distributed reconfigurable intelligent surface, semantic communication, industrial Internet-of-Things, joint communication and computation design.
\end{IEEEkeywords}

\section{Introduction}
Reconfigurable intelligent surfaces (RISs) exhibit great potential in providing feasible and energy-efficient solutions to massive data traffic and high reliability of wireless communications at low cost \cite{9475160,9899454,9530717}. Specifically, an RIS is usually comprised of a large number of hardware-efficient and nearly passive reflecting elements, each of which can alter the phase of the incoming signal, without requiring a dedicated power amplifier \cite{ huang2019reconfigurable,9140329,gan2024bayesian}. Thus, the deployment of RISs enables the manipulation of electromagnetic waves, and brings benefits in low-power, energy-efficient, high-speed, massive connectivity, and low-latency wireless communications \cite{9206044,9110849,9779586}.

Inspired by these advantages, distributed RIS structure is further proposed to serve more users and complicated scattering environment for additional profits, such as enlarged communication coverage, enhanced spectral efficiency and more flexible design \cite{9746287,9482600}. For example, \cite{9746287} showed that double RIS systems improved channel estimation accuracy at less training overhead compared with the single RIS systems, since the fewer channel coefficients are required in careful element distribution design in double RISs. \cite{9501034} proposed that increasing the number of RISs can improve the system performance even with timing synchronization errors.

However, the joint beamforing design of multi-RIS systems is much more complicated due to the existence of a large amount of RIS elements and cooperative communications between different RISs. For example, \cite{9241706} iteratively optimized beamforming vectors for different RISs in double RIS systems using traditional eigenmode beamforming schemes.  Besides, the work in \cite{9105111} designed beamforming schemes in an iterative behavior using Lagrangian method and Riemannian manifold conjugate gradient method, showing that multi-RIS design exhibited higher performance gain than single RIS systems. Furthermore, the authors in \cite{9241752} adopted graph-optimization algorithms to design beam routing in multi-RIS systems, and showed a trade-off between path loss minimization and cooperative beamforming gain maximization.

Nevertheless, the most existing beamforming designs in multi-RIS wireless communications require complex optimization process in an iterative manner, and the complexity dramatically increase with the number of RISs, which hinders practical adoption in multi-RIS systems. Therefore, we propose an efficient beamforming design in multi-RIS systems in this work. Specifically, we adopt the graph theory based algorithm to solve the RIS-user association problem and the tensor based scheme is applied to solve the RIS phase optimization. 

{\color{black} Semantic communication has recently garnered significant attention in the field of wireless communications \cite{gunduz2022beyond,xu2022edge,guler2018semantic,qin2021semantic,luo2022semantic}. This emerging paradigm involves joint source and channel coding \cite{dai2022nonlinear}, wireless resource allocation \cite{zhao2024prob,xia2022wireless}, learning resource allocation \cite{tong2021federated2}, etc. In contrast to conventional communication, which is bit-centric, semantic communication focuses on transmitting the underlying meaning of the message \cite{lu2022rethinking,10233741,chen2022performance}. The semantic information of a message can be extracted using advanced artificial intelligence (AI) techniques and can be tailored to the receiver's specific task \cite{xie2021deep,wang2021performance}. It is important to note that the process of extracting semantic information requires additional computational resources \cite{10032275,zhao2024joint}.

One of the key advantages of semantic communication is its efficiency. The data size of the meaning of a message is typically much smaller than the message itself, resulting in reduced communication latency compared to conventional communication schemes \cite{10287956,10149174}. Furthermore, semantic communication is characterized by its reliability. In conventional communication, which relies on bits, a small bit error can lead to a significant error in meaning. Conversely, in semantic communication, the transmitted information remains similar even if there is a small error during the decoding process, as it operates on the semantic dimension \cite{9953079,10061867,10101778}.

Due to its efficiency and reliability, semantic communication is envisioned to be a key technology for realizing future industrial Internet-of-Things (IIoT) networks. By providing seamless and reliable transmission, semantic communication has the potential to revolutionize the way data is exchanged and processed in IIoT environments. As industries continue to embrace digital transformation and the integration of connected devices, the adoption of semantic communication techniques is expected to play a crucial role in enabling efficient and robust communication within these complex systems.

In recent years, several studies have explored semantic communication in the context of IIoT. For example,
the authors of \cite{10405134} introduced a semantic-driven cross-modal obstacle detection scheme which jointly utilizes semantic encoding for real-time transmission and enhances detection precision by exploring and exploiting correlations among heterogeneous modalities. In \cite{9170818}, the authors investigated the collaborative deep neural network inference problem in IIoT networks. They formulated the problem as a constrained Markov decision process, accounting for channel variation and task arrival randomness, and proposed a deep reinforcement learning-based algorithm to solve it. Furthermore, a lightweight semantic knowledge base (SKB)-enabled multi-level feature extractor was developed in \cite{10287247}. This innovative approach projects images into visual, semantic, and intermediate feature spaces. The authors also proposed an SKB-enabled multi-level feature transmission framework that utilizes the SKB and multi-level feature extractor at both the transmitter and receiver.
However, a critical aspect that these studies have neglected is the semantic computation consumption. This factor is crucial in the joint communication and computation design for semantic communication, as the transmitter is responsible for both the extraction of semantic information through computation and the transmission of wireless information.}

Combining the advantages of both distributed RISs and semantic communication, this paper investigates a joint communication and computation framework for distributed RISs assisted probabilistic semantic communication (PSC). The main contributions of this paper are summarized as follows:
\begin{itemize}
    \item We propose a joint communication and computation framework for distributed RISs assisted PSC. In the considered model, distributed RISs are deployed to serve multiple users. For semantic communication, the PSC model which facilitates semantic communication through shared probability graphs between the base station (BS) and users is considered, and the compute-then-transmit protocol is proposed. Considering both semantic compression and computation factors, we formulate the semantic-aware transmission rate.
    \item For the considered model, the semantic-aware sum rate maximization problem is formulated with considering transmit power, RIS-user association, phase shift, and semantic compression ratio constraints. To solve the formulated mixed integer optimization problem, a many-to-many matching algorithm is proposed to solve the RIS-user association subproblem, where a stable matching can be obtained with low complexity. Besides, the semantic compression ratio subproblem is tackled using greedy algorithm.
    \item For the phase shift optimization of one RIS serving multiple users, we consider four kinds of multiple access, transmit beamforming multiple access, surface space division multiple access, frequency division multiple access, and time division multiple access. For the phase shift optimization of multiple RISs serving one user, the optimal phase design is obtained in closed form.
    \item Simulation results verify the superiority of the proposed scheme compared to the conventional schemes.
\end{itemize}

The remainder of this paper is organized as follows. The system model and problem formulation are described in Section \ref{section2}. The algorithm design is presented in Section \ref{section3}. Simulation results are analyzed in Section \ref{section4}. Conclusions are drawn in Section \ref{section5}.

\section{System Model and Problem Formulation}\label{section2}
\begin{figure}[t]{\color{black} 
    \centering
    \includegraphics[width=\linewidth]{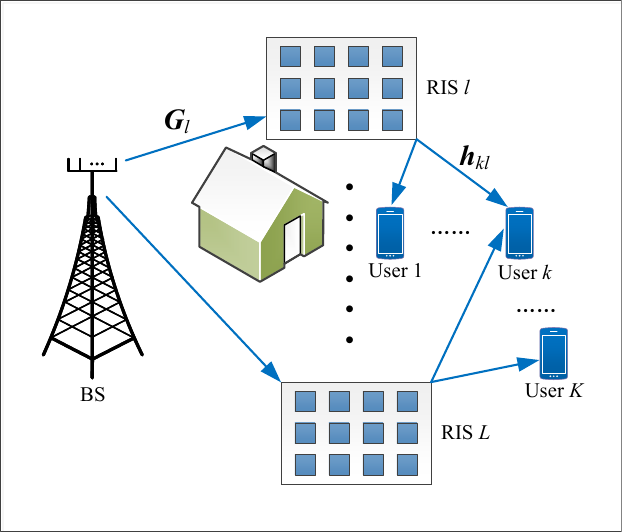}
    \caption{A wireless communication system with distributed RISs, one BS, and multiple users.}
    \label{fig1}}
\end{figure}

Consider a single-cell network with multiple distributed RISs and one multi-antenna BS serving several single-antenna users, as shown in Fig.~\ref{fig1}. The number of RISs and users are respectively denoted by $L$ and $K$. The number of antennas at the BS is denoted by $N$. Without loss of generality, the number of elements for each RIS is the same with $N_0$.

\subsection{Channel Model}
The BS is equipped with uniform linear array (ULA), while RISs utilize uniform planar array (UPA). {\color{black} For ULA, the normalized array response vector can be given by
\begin{equation}\label{sys1eq1}
    \boldsymbol{a}\left(\phi, N\right) = \frac{1}{{\sqrt{N} }}{\left[1, \ldots, e^{j\pi (N-1) \phi}\right]^{\mathrm{T}}}.
\end{equation}}
Thus, the array response vector of the BS can be given by
\begin{equation}\label{sys1eq2}
    \boldsymbol{a}_B(\varphi) = \boldsymbol{a}\left(\frac{{2\pi d_B}}{\lambda} \sin \left(\varphi\right), N\right),
\end{equation}
where $d_B$ is the distance between two adjacent antennas of the BS, $\lambda$ is the signal wavelength, and $\varphi$ is the incidence angle. 

For UPA, the normalized array response vector can be given by
\begin{align}\label{sys1eq3}
    \boldsymbol{a}_I(\varphi^a, \varphi^e) =& \boldsymbol{a}\left(\frac{{2\pi d_I}}{\lambda} \sin \left(\varphi^e\right)\cos \left(\varphi^a\right), M_1\right)
    \nonumber\\
    &\otimes \boldsymbol{a}\left(\frac{{2\pi d_I}}{\lambda } \cos \left(\varphi^e\right), M_2\right),
\end{align}
where $\otimes$ means Kronecker product, $d_I$ is the distance between two adjacent antennas of the RIS, $M_1$ is the number of elements in the horizontal direction, $M_2$ is the number of elements in the vertical direction, $\varphi^a$ and $\varphi^e$ are the angles in the horizontal and vertical directions, respectively. In our case, $N_0=M_1 M_2$.

With \eqref{sys1eq2} and \eqref{sys1eq3}, the channel gain from the BS to RIS $l$ is
\begin{align}
    \boldsymbol G_l=\boldsymbol{a}_I(\varphi^a_{l}, \varphi^e_{l}) \boldsymbol{a}_B^\mathrm{H}(\varphi_{l}),
\end{align}
where $\varphi^a_{l}$ and $\varphi^e_{l}$ are respectively the azimuth and elevation angles of arrival of RIS $l$, and $\varphi_{l}$ is the azimuth angle of departure of the BS. 




The channel gain between RIS $l$ and user $k$ can be expressed as
\begin{align}
    \boldsymbol h_{kl}=\boldsymbol{a}_I(\varphi^a_{lk}, \varphi^e_{lk}),
\end{align}
where $\varphi^a_{lk}$ and $\varphi^e_{lk}$ are respectively the azimuth and elevation angles of departure of RIS $l$.

\subsection{Transmission Model}
In this paper, since there are multiple RISs distributed in the whole space, we consider the RIS-user association problem. 
Let $x_{kl}\in\{0,1\}$ denote the association between user $k$ and RIS $l$.
Assume that the direct signal between the BS and all users are blocked due to obstacles and high buildings. 
Considering the signals reflected from multiple RISs, the received signal at user $k$ is
\begin{equation}\label{sys2eq2}
    y_k=\sum_{l=1}^L x_{kl}\boldsymbol h_{kl}^\mathrm{H}\boldsymbol\Theta_l  \boldsymbol G_l\boldsymbol s+n_k,
\end{equation}
where $\boldsymbol \Theta_l=\mathrm{diag} \left(e^{j\theta_{l1}}, \cdots, e^{j\theta_{lN_0}}\right)\in\mathbb C^{N_0\times  N_0}$ with $\theta_{ln}\in[0,2\pi]$, $l\in\mathcal L=\{1, 2, \cdots, L\}$, $n\in\mathcal N_0=\{1, 2, \cdots, N_0\}$, and $n_k\sim\mathcal{CN}(0,\sigma^2)$ is the additive white Gaussian noise, in which $\mathrm{diag}\left(e^{j\theta_{l1}}, \cdots, e^{j\theta_{lN_0}}\right)$ is a diagonal matrix with $\left\{e^{j\theta_{l1}}, \cdots, e^{j\theta_{lN_0}}\right\}$ being its diagonal elements.

Based on \eqref{sys2eq2}, the achievable rate of user $k$ can be given by
\begin{equation}\label{sys2eq3_1}
    r_{k}=B \log_2\left(1+\frac{ p_k\left| 
        \sum_{l=1}^L x_{kl}\boldsymbol h_{kl}^\mathrm{H} \boldsymbol\Theta_l  \boldsymbol G_l \boldsymbol w_k\right|^2}
    {\sum_{i=1,i\neq k}^K  p_i \left| 
        \sum_{l=1}^L  \boldsymbol h_{kl}^\mathrm{H} \boldsymbol\Theta_l \boldsymbol G_l \boldsymbol w_i\right|^2+\sigma^2}\right),
\end{equation}
where $B$ is the bandwidth of the system, $p_k$ is the allocated power of user $k$, and $\boldsymbol w_k$ is the beamforming vector of user $k$ with satisfying $\|\boldsymbol w_k\|=1$.

\subsection{Semantic Computation and Transmission Models}
\begin{figure*}[t]
    \centering
    \includegraphics[width=\linewidth]{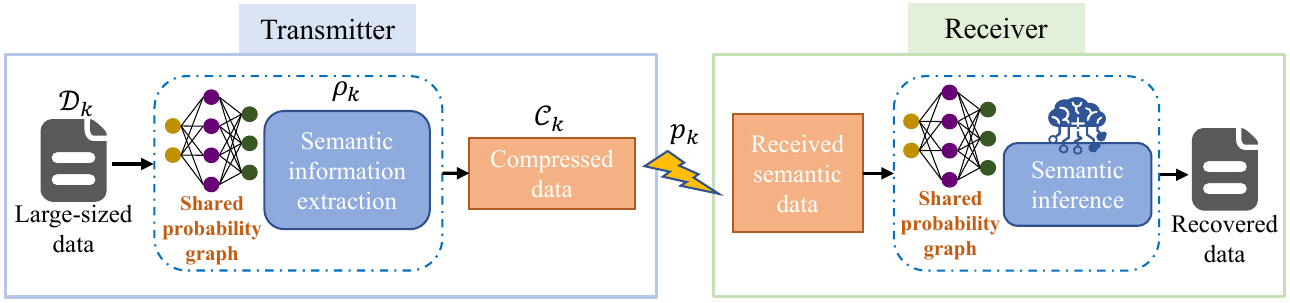}
    \caption{The compute-then-transmit protocol for probabilistic semantic communication systems.}
    \label{fig2}
\end{figure*}
In this paper, the PSC model is considered, which adopts the compute-then-transmit protocol \cite{10333452} as shown in Fig.~\ref{fig2}. The original information intended for each user is first computed to extract the small-sized semantic information, which be then transmitted to the user.

Within the PSC network, each user shares its individualized probability graph with the BS. Using these shared probability graphs, the BS performs semantic information extraction, compressing the large-sized data $\mathcal{D}_k$ for each user based on their respective probability graphs with semantic compression ratio denoted by $\rho_k$. Subsequently, the compressed data $\mathcal{C}_k$ is transmitted to user $k$. When receiving the semantic data from the BS, each user performs semantic inference, using its local probability graph to restore the compressed semantic information.

Mathematically, the semantic compression ratio for user $k$ can be written as
\begin{equation}
    \rho_k=\frac{\mathrm{size}(\mathcal{C}_k)}{\mathrm{size}(\mathcal{D}_k)},
\end{equation}
where the function $\mathrm{size}(\cdot)$ quantifies the data size in terms of bits.

Consequently, we can obtain a semantic-aware transmission rate for user $k$, denoted by
\begin{equation}
    c_k = \frac{1}{\rho_k}r_k,
\end{equation}
which represents the actual transmission rate of the semantic information. The conventional transmission rate is multiplied by $1/\rho_k$ because one bit of extracted semantic information can carry more than one bit of the original information.

In the considered PSC network, the BS executes semantic information extraction. This process is closely linked to computational resources, with a crucial observation that lower semantic compression ratio corresponds to increased computation overhead.

{\color{black} According to equations (18)-(20) in \cite{zhao2023joint}, the computation overhead of user $k$ in PSC network can be described by
\begin{equation}\label{cl}
    f_k\left(\rho_k\right)=\begin{cases}
        A_{k1}\rho_k +B_{k1}, &C_{k1}< \rho_k \leq C_{k0}, \\
        A_{k2}\rho_k +B_{k2}, &C_{k2}< \rho_k \leq C_{k1}, \\
        \vdots \\
        A_{kD}\rho_k +B_{kD}, &C_{kD}\leq \rho_k \leq C_{k(D-1)},
    \end{cases}
\end{equation}
where $A_{kd} < 0$ denotes the slope, $B_{kd} > 0$ represents constant term ensuring the continue connection of each segment, $C_{kd}$ is the boundary for each segment, and $C_{k0}=1$. These parameters are closely related to the characteristics of the probability graphs involved.}

\begin{figure}[t]
    \centering
    \includegraphics[width=\linewidth]{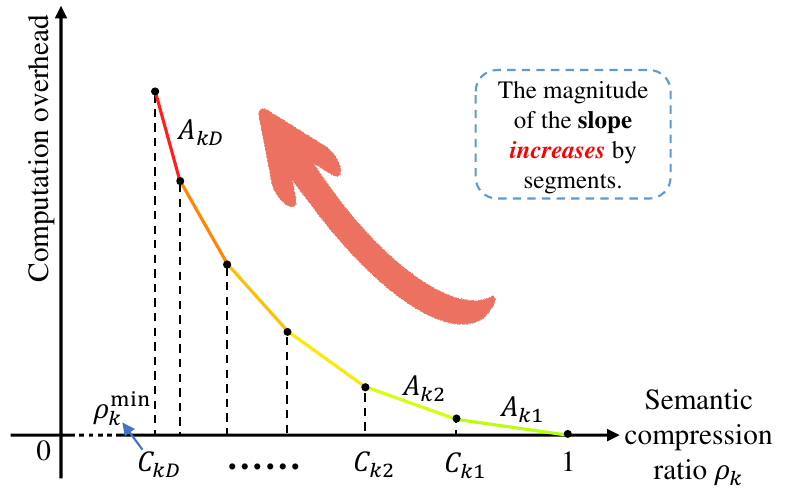}
    \caption{Illustration of the computation overhead versus the semantic compression ratio.}
    \label{fg:cl}
\end{figure}

As shown in Fig.~\ref{fg:cl}, the computation overhead function features a segmented architecture with $D$ levels, and the magnitude of the slope grows in segments with lower semantic compression ratio. This is due to the fact that at higher semantic compression ratios, only low-dimensional conditional probabilities are used, resulting in lower computational requirements. However, when the semantic compression ratio decreases and surpasses certain thresholds, such as $C_{kd}$, the need for higher-dimensional information arises, which results in increased computation overhead. Each transition in the segmented function $f_k(\rho_k)$ indicates the inclusion of probabilistic information with a higher dimension for semantic information extraction. Notably, achieving a higher semantic-aware transmission rate requires a lower semantic compression ratio, which in turn requires a higher computation overhead.

\subsection{Problem Formulation}
Based on the considered model, our aim is to maximize the sum semantic-aware transmission rate of the whole system through optimizing the RIS-user association, transmit power, beamforming vector, phase shift, and semantic compression ratio of each user. This joint communication and computation optimization problem can be mathematically formulated as
\begin{subequations}\label{sys2max0}
    \begin{align}
        \max_{\boldsymbol x, \boldsymbol p, \boldsymbol w, \boldsymbol \theta, \boldsymbol \rho} & \sum_{k=1}^K \frac{B} {\rho_k} \log_2\Bigg(1+ \notag \\
        & \qquad \frac{p_k\left| \sum_{l=1}^L x_{kl}\boldsymbol h_{kl}^\mathrm{H} \boldsymbol\Theta_l  \boldsymbol G_l \boldsymbol w_k\right|^2}{\sum_{i=1,i\neq k}^K  p_i \left| \sum_{l=1}^L  \boldsymbol h_{kl}^\mathrm{H} \boldsymbol\Theta_l \boldsymbol G_l \boldsymbol w_i\right|^2+\sigma^2}\Bigg), \tag{\ref{sys2max0}}\\
	\textrm{s.t.} \quad
		& \sum_{k=1}^K x_{kl}\leq K_0, \quad \forall l\in\mathcal{L},
		\\& \sum_{l=1}^L x_{kl}\leq L_0, \quad \forall k\in\mathcal{K},
		\\& x_{kl}\in\{0,1\}, \quad \forall k\in\mathcal{K}, \forall l\in\mathcal{L},
        \\& \sum_{k=1}^K p_{k}\leq P,
        \\& p_k\geq 0, \quad \forall k\in\mathcal{K},
        \\& \rho_k^{\min}\leq\rho_k\leq 1, \quad \forall k\in\mathcal{K},
        \\& \sum_{k=1}^K f_k(\rho_k)\leq Q,
		\\& \|\boldsymbol w_k\|=1, \quad \forall k\in\mathcal{K},
		\\& \theta_{ln}\in[0,2\pi],  \quad \forall l\in\mathcal{L}, \forall n\in\mathcal{N}_0,
    \end{align}
\end{subequations}
where $\boldsymbol \theta=[\theta_{11},\cdots,\theta_{1N_0},\cdots, \theta_{LN_0}]^\mathrm{T}$, $\boldsymbol w=[\boldsymbol w_1; \cdots; \boldsymbol w_K]$, $\boldsymbol x=[x_{11},\cdots,x_{1L},\cdots,x_{KL}]^\mathrm{T}$, $\boldsymbol \rho=[\rho_{1},\cdots,\rho_{K}]^\mathrm{T}$, $\mathcal{K}=\{1,2,\cdots,K\}$ denotes the set of users, $K_0$ is the maximum number of associated users for each RIS, $L_0$ is the maximum number of associated RISs for each user, $P$ is the maximum transmit power of the BS, $Q$ is the maximum computation budget of the BS, and $\rho_k^{\min}$ is the minimum semantic compression ratio of user $k$.

{\color{black} It is generally challenging to solve problem \eqref{sys2max0} as it is a mixed integer optimization problem, which typically incurs exponential computation complexity to obtain the optimal solution. Moreover, the objective function is non-convex, and the presence of the piecewise function $f_k(\rho_k)$ further complicates the optimization process. To address these challenges and achieve a polynomial-time solution for the problem \eqref{sys2max0}, we propose an iterative algorithm that leverages the alternating method in the next section.}

\section{Algorithm Design}\label{section3}
To solve the complicated mixed integer optimization problem \eqref{sys2max0}, we first provide the many-to-many matching algorithm to solve the RIS-user association subproblem, then obtain the optimized semantic compression ratio under greedy policy, the phase shift optimization is based on tensor beamforming, and the optimal power control is obtained by solving a convex problem. {\color{black} These four subproblems are iteratively optimized until convergence.}

\subsection{Many-to-Many Matching}
To solve the RIS-user association subproblem, we utilize the many-to-many matching to solve the association vector $\boldsymbol x$ optimization.

{\color{black} To perform the many-to-many matching, we first sort the channel gains of each user reflected by each RIS in descending order, which is stored in a preference list. The preference list of user $k$ is denoted by $\mathcal S_k$, and the index of the most preferred RIS of user $k$ is denoted by $l_k$. Denote the number of RISs with which user $k$ is associated by $L_k$, and the number of users that RIS $l$ is associated with by $K_l$. Then, for each user, we conduct the following procedure.

If the intended RIS $l_k$ is not occupied and the maximum number of associated RISs for user $k$ does not exceed $L_0-1$, i.e., $L_k\leq L_0-1$, associate RIS $l_k$ with user $k$. Delete index $l_k$ from the preference list $\mathcal{S}_k$.

If the intended RIS $l_k$ is occupied, the maximum number of associated RISs for user $k$ does not exceed $L_0-1$, and the maximum number of associated users for RIS $l_k$ does not exceed $K_0-1$, associate user $k$ with RIS $l_k$. Delete index $l_k$ from the preference list $\mathcal{S}_k$.

If the intended RIS $l_k$ is occupied, the maximum number of associated RISs for user $k$ does not exceed $L_0-1$, and the maximum number of associated users for RIS $l$ is $K_0$, associate user $k$ with RIS $l_k$ if the objective function will increase with one user associated with RIS $l_k$ replaced by user $k$. Update the associated set of RIS $l_k$ be replacing the user with user $k$, which can mostly increase the objective value. Delete index $l_k$ from the preference list $\mathcal{S}_k$.

This procedure terminates when $\mathcal S_k = \emptyset, \forall k \in \mathcal K$.

The many-to-many matching for RIS-user association is summarized in Algorithm \ref{multipleusergreedy}.}

\begin{algorithm}[t]
    \caption{Many-to-Many Matching for RIS-User Association}{\color{black}
    \begin{algorithmic}[1]\label{multipleusergreedy}
        \STATE Initialize $L_k=0,\forall k\in\mathcal K$ and $K_l=0,\forall l\in\mathcal L$.
        \WHILE{$\mathcal S_k \neq \emptyset, \exists k \in \mathcal K$}
            \FOR{$k=1:K$}
                \IF{$L_k<L_0$}
                    \IF{$K_{l_k}<K_0-1$}
                        \STATE Associate RIS $l_k$ with user $k$.
                    \ENDIF    
                    \IF{$K_{l_k}=K_0$ and the objective function increases with one user associated with RIS $l_k$ replaced by user $k$}
                        \STATE Associate RIS $l_k$ with user $k$.
                        \STATE Update the associated set of RIS $l_k$ be replacing the user with user $k$, which can mostly increase the objective value.
                    \ENDIF
                \ENDIF
                \STATE Delete index $l_k$ from the preference list $\mathcal{S}_k$.
                \STATE Update $L_k$ and $K_l$.
            \ENDFOR
        \ENDWHILE
        \STATE \textbf{Output:} The optimized RIS-user association vector $\boldsymbol x$.
    \end{algorithmic}}
\end{algorithm}

\subsection{Semantic Compression Ratio Optimization}
According to problem \eqref{sys2max0}, the semantic compression ration optimization problem can be given by
\begin{subequations}\label{Al3Sema1}
    \begin{align}
        \max_{\boldsymbol \rho} \quad &  \sum_{k=1}^K \frac{1}{\rho_k} r_k,\tag{\ref{Al3Sema1}}\\
        \textrm{s.t.} \quad
            & \rho_k^{\min}\leq\rho_k\leq 1, \quad \forall k\in\mathcal{K},
            \\& \sum_{k=1}^K f_k(\rho_k)\leq Q.\label{q}
    \end{align}
\end{subequations}
The difficulty in solving problem \eqref{Al3Sema1} lies in the piecewise function $f_k(\rho_k)$. To address this difficulty, we propose the use of binary variable $\theta_{kd}\in \{0,1\}$ to indicate the linear segment level of user $k$. When $\theta_{kd} =1$, the computation overhead function $f_k(\rho_k)$ belongs to the $d$-th segment for user $k$, which can be expressed as $f_k(\rho_k)=A_{kd}\rho_k+B_{kd}$. On the other hand, when $\theta_{kd}=0$, the computation overhead function does not fall within the $d$-th segment. By introducing the binary variable $\theta_{kd}$, we can reformulate the computation overhead function as
\begin{equation}
    f_k(\rho_k)=\sum_{d=1}^{D} \theta_{kd}(A_{kd}\rho_k+B_{kd}),
\end{equation}
where $D$ is the total number of segments of the piecewise function, and $\theta_{kd}$ identifies the specific segment that $\rho_k$ belongs to. Given that each user possesses only a single linear segment level, we have
\begin{equation}
    \sum_{d=1}^{D}\theta_{kd}=1,\theta_{kd}\in \{0,1\}.
\end{equation}

Based on problem \eqref{Al3Sema1}, it is obvious that the lower the semantic compression ratio, the higher the semantic-aware transmission rate. Therefore, we should set the semantic compression ratio of each user as low as possible while satisfying constraint \eqref{q}. From Fig.~\ref{fg:cl}, we can see that when semantic compression ratio is high, a small computation overhead can significantly reduce the ratio. However, when semantic compression ratio is low, a large computation overhead can only slightly reduce it. Hence, under limited computation resource budget, we should first guarantee that the semantic compression ratio of every user are all at a relatively low level which is more efficient.

First, we assume the semantic compression ratio of all users are at the same segment, which is
\begin{align}
    & \theta_{1d}=\theta_{2d}=\cdots=\theta_{Kd}=1,\\
    & \theta_{1j}=\theta_{2j}=\cdots=\theta_{Kj}=0,\forall j\neq d.
\end{align}
Then, we assume the semantic compression ratio of all users are at the smallest value at segment $d$, which is
\begin{equation}
    \rho_1=C_{1(d-1)},\rho_2=C_{2(d-1)},\cdots,\rho_K=C_{K(d-1)}.
\end{equation}
Thus, the total computation overhead at segment $d$ can be calculated as
\begin{equation}
    Q_d=\sum_{k=1}^K f_k(\rho_k)=\sum_{k=1}^K A_{kd}C_{k(d-1)}+B_{kd}.
\end{equation}

Apparently, when $d$ increases, $Q_d$ grows, i.e.,
\begin{equation}
    Q_1 < Q_2 < \cdots < Q_D,
\end{equation}
where $Q_1=0$ since when $d=1$, $\rho_k=C_{k0}=1,\forall k\in\mathcal{K}$. Denote
\begin{equation}
    Q_{D+1}=\sum_{k=1}^K f_k(\rho_k^{\min})=\sum_{k=1}^K A_{kD}C_{kD}+B_{kD}.
\end{equation}
If $Q\geq Q_{D+1}$, we can simply set $\rho_k=\rho_k^{\min},\forall k\in\mathcal{K}$ for the maximum sum semantic-aware transmission rate. Otherwise, if $Q_{d^*}\leq Q<Q_{d^*+1}$, we first set
\begin{equation}
    \rho_k=C_{k(d^*-1)},\forall k\in\mathcal{K}.
\end{equation}
Then, following the greedy policy, users with larger $r_k$ should be assigned a lower semantic compression ratio since they contribute more to the sum semantic-aware transmission rate.

{\color{black} Without loss of generality, we assume that user $1$ has the highest $r_k$, and user $K$ has the lowest $r_k$, respectively, that is,}
\begin{equation}
    r_1 \geq r_2 \geq \cdots \geq r_K.
\end{equation}
When the computation budget of the BS is sufficient, we set the semantic compression ratio of the user corresponding to the currently highest $r_k$ to the lowest value of segment $d^*$, i.e., $C_{kd^*}$. Denote the available additional computation overhead for user $k$ by
\begin{equation}
    Q^{(k)}=Q-Q_{d^*}-\sum_{i=1}^{k-1} A_{id^*}\left(C_{id^*}-C_{i(d^*-1)}\right),k>1,
\end{equation}
and $Q^{(1)}=Q-Q_{d^*}$.
Obviously, $Q^{(k)}$ decreases when $k$ increases, i.e.,
\begin{equation}
    Q^{(1)}\geq Q^{(2)}\geq\cdots\geq Q^{(K)}.
\end{equation}
If $Q^{(k)}\geq A_{kd^*}\left(C_{kd^*}-C_{k(d^*-1)}\right)$, we set $\rho_k=C_{kd^*}$. If $A_{kd^*}\left(C_{kd^*}-C_{k(d^*-1)}\right)>Q^{(k)}>0$, we can calculate $\rho_k=C_{k(d^*-1)}+Q^{(k)}/A_{kd^*}$. Otherwise, if $Q^{(k)}\leq 0$, we set $\rho_k=C_{k(d^*-1)}$. Therefore, the optimized semantic compression ratio of user $k$ can be given by
\begin{equation}\label{scr}
    \rho_k=\left\{\begin{array}{l}
        C_{kd^*}, \hspace{2.7em}\textrm{if } Q^{(k)}\geq A_{kd^*}\left(C_{kd^*}-C_{k(d^*-1)}\right), \\
        C_{k(d^*-1)}, \quad\textrm{if } Q^{(k)}\leq 0,\\
        C_{k(d^*-1)}+\frac{Q^{(k)}}{A_{kd^*}}, \quad\textrm{otherwise}.
    \end{array}\right..
\end{equation}



\subsection{Transmit and Phase Beamforming Design}\label{3c}
We consider the following two cases for RIS-user association: one RIS serves multiple users and multiple RISs serve one user.

\subsubsection{One RIS Servers Multiple Users}
{\color{black} Take RIS $l$ as an example, the set of users associated with RIS $l$ is denoted by $\mathcal U_l$. To mitigate the multi-user interference, the zero forcing (ZF) beamforming technique is adopted in this work. Since multiple users are served by the same RIS, we consider the following four types of multiple access schemes.}

$\bullet$\quad\textbf{\emph{Non-orthogonal Multiple Access (NOMA)}}

In the NOMA scheme, the whole RIS $l$ is occupied by all served users simultaneously on the same bandwidth.  	 
Mathematically, the joint transmit and passive beamforming optimization problem can be formulated as
\begin{subequations}\label{zfmax1}
    \begin{align}
        \max_{\boldsymbol\theta_l, \boldsymbol w_k} \quad&
        \left|\boldsymbol h_{kl}^\mathrm{H} \mathrm{diag}(\boldsymbol \theta_l) \boldsymbol G_l \boldsymbol w_k\right| \tag{\theequation}\\
        \textrm{s.t.} \quad
        &\boldsymbol h_{il}^\mathrm{H} \boldsymbol\Theta_l  \boldsymbol G_l  \boldsymbol w_k=0,\quad \forall i\neq k,i\in\mathcal U_l,\\
        &\theta_{ln} \in[0,2\pi],\quad \forall n\in\mathcal{N}_0,\\
        &\|\boldsymbol w_k\|=1,
    \end{align}
\end{subequations}
where $\boldsymbol \theta_l=[\theta_{l1},\cdots, \theta_{lN_0}]$.
	
Problem \eqref{zfmax1} can be solve in two steps, i.e, phase shift optimization in the first step, and transmit beamforming in the second step.
In the first step, we optimize the phase shift $\boldsymbol \theta_l$. 
The objective function in \eqref{zfmax1} can be written as
\begin{align}\label{zfmax1eq1}
    &\boldsymbol h_{kl}^\mathrm{H} \mathrm{diag}(\boldsymbol \theta_l) \boldsymbol G_l
    \nonumber\\=&\boldsymbol{a}_I^\mathrm{H}(\varphi^a_{lk}, \varphi^e_{lk})  \mathrm{diag}(\boldsymbol \theta_l) \boldsymbol{a}_I(\varphi^a_l, \varphi^e_l) \boldsymbol{a}_B^\mathrm{H}(\varphi_l)
    \nonumber\\=&
    \boldsymbol{a}_I^\mathrm{H}(\varphi^a_{lk}, \varphi^e_{lk})  \odot \boldsymbol{a}_I^\mathrm{T}(\varphi^a_l, \varphi^e_l)
    \boldsymbol \theta_l \boldsymbol{a}_B^\mathrm{H}(\varphi_l)
    \nonumber\\=&
    \boldsymbol{a}^\mathrm{H}\left(\frac{{2\pi d_I}}{\lambda} (\sin \left(\varphi^e_{lk}\right)\cos \left(\varphi^a_{lk}\right)
    -\sin \left(\varphi^e_{l}\right)\cos \left(\varphi^a_{l}\right)
    ), M_1\right)
    \nonumber\\ &
    \otimes \boldsymbol{a}^\mathrm{H}\left(\frac{{2\pi d_I}}{\lambda } (\cos \left(\varphi^e_{lk}\right)-\cos \left(\varphi^e_{l}\right)), M_2\right)
    \boldsymbol \theta_l \boldsymbol{a}_B^\mathrm{H}(\varphi_l),
\end{align}
and
\begin{equation}
    \boldsymbol w_k=\frac{\boldsymbol{a}_B^\mathrm{H}(\varphi_l)}{|\boldsymbol{a}_B^\mathrm{H}(\varphi_l)|},
\end{equation}
where $\odot$ means Hadamard product.

{\color{black} To maximize the effective channel gain for user $k$,  the optimal phase design can be derived from \eqref{zfmax1eq1}, i.e., 
\begin{align}\label{zfmax1eq2}
    \boldsymbol \theta_l^k=&
    \boldsymbol{a}\left(\frac{{2\pi d_I}}{\lambda } (\sin \left(\varphi^e_{lk}\right)\cos \left(\varphi^a_{lk}\right)
    -\sin \left(\varphi^e_{l}\right)\cos \left(\varphi^a_{l}\right)
    ), M_1\right)
    \nonumber\\ &
    \otimes \boldsymbol{a}\left(\frac{{2\pi d_I}}{\lambda } (\cos \left(\varphi^e_{lk}\right)-\cos \left(\varphi^e_{l}\right)), M_2\right).
\end{align}

Note that RIS $l$ needs to serve multiple users at the same time, the sum rate of all users served by RIS $l$ can be approximated by}
\begin{align}\label{zfmax1eq3}
    &\sum_{k\in\mathcal U_l}B  \log_2\left(1+\frac{p_k\left| 
    \boldsymbol h_{kl}^\mathrm{H} \boldsymbol\Theta_l  \boldsymbol G_l \boldsymbol w_k\right|^2}
    {\sigma^2}\right)
    \nonumber\\ \approx& \sum_{k\in\mathcal U_l}B \log_2\left(\frac{ p_k\left| 
    \boldsymbol h_{kl}^\mathrm{H} \boldsymbol\Theta_l  \boldsymbol G_l \boldsymbol w_k\right|^2}
    {\sigma^2}\right)
    \nonumber\\ =&B  \log_2\left(\frac{\prod_{k\in\mathcal U_l}p_k\left| 
    \boldsymbol h_{kl}^\mathrm{H} \boldsymbol\Theta_l  \boldsymbol G_l \boldsymbol w_k\right|^2}
    {\sigma^2}\right)
    \nonumber\\ \approx&B  \log_2\left(\frac{\left| 
    \hat{\boldsymbol h}_{l}^\mathrm{H} \boldsymbol (\Theta_l  \boldsymbol G_l)^{|\mathcal U_l|}  \right|^2\prod_{k\in\mathcal U_l}p_k}
    {\sigma^2}\right),
\end{align}
with 
\begin{equation}
    \hat{\boldsymbol h}_{l} = {\boldsymbol h}_{lk_1} \odot \cdots \odot {\boldsymbol h}_{lk_{|\mathcal U_l|}},
\end{equation}
where $\mathcal U_l=\{k_1,\cdots, k_{|\mathcal U_l|}\}$. 
In the last approximation of \eqref{zfmax1eq3}, we use $\|\boldsymbol w_k\|=1$.
{\color{black} Based on \eqref{zfmax1eq2} and \eqref{zfmax1eq3}, the phase shift of RIS $l$ can be formulated by}
\begin{equation}\label{case1ZFeq0}
    \theta_{ln}^*=\frac{\sum_{k\in\mathcal U_l} \theta_{ln}^k}{|\mathcal U_l|}.
\end{equation}

{\color{black} For the transmit beamforming at the BS, $\boldsymbol w_k$ is the same for all served users. In this case, users can be served by using power-domain NOMA.}
 
$\bullet$\quad\textbf{\emph{Surface Space Division}}
	
In surface space division scheme, RIS $l$ is uniformly divided into multiple small parts and each part is occupied by one user. Mathematically, the joint transmit and passive beamforming optimization problem can be formulated as
\begin{subequations}\label{case2zfmax1}
    \begin{align}
        \max_{\boldsymbol\theta_{kl}, \boldsymbol w_k} \quad&
        \left|\boldsymbol h_{kl}^\mathrm{H} \mathrm{diag}(\boldsymbol \theta_{kl})   \boldsymbol G_{kl} \boldsymbol w_k\right| \tag{\theequation}\\
        \textrm{s.t.} \quad
        &\boldsymbol h_{il}^\mathrm{H} \boldsymbol\Theta_{il} \boldsymbol G_{il} \boldsymbol w_k=0, \quad \forall i\neq k,i\in\mathcal U_l,\\
        &\theta_{ln} \in[0,2\pi], \quad \forall n\in\mathcal{N}_0,\\
        &\|\boldsymbol w_k\|=1,
    \end{align}
\end{subequations}
where $\boldsymbol \theta_{kl}$ is the phase beamforming of the part in RIS $l$ allocated to user $k$.

In this case, we have 
\begin{align}
    \boldsymbol \theta_{kl}^*=&
    \boldsymbol{a}\left(\frac{{2\pi d_I}}{\lambda } (\sin \left(\varphi^e_{lk}\right)\cos \left(\varphi^a_{lk}\right)
    -\sin \left(\varphi^e_{l}\right)\cos \left(\varphi^a_{l}\right)
    ), \hat M_1\right)
    \nonumber\\ &
    \otimes \boldsymbol{a}\left(\frac{{2\pi d_I}}{\lambda } (\cos \left(\varphi^e_{lk}\right)-\cos \left(\varphi^e_{l}\right)), \hat M_2\right),
\end{align}
where $\hat M_1$ and $\hat M_2$ are the number of elements in the horizontal and vertical directions allocated for user $k$.

Given $\boldsymbol\theta_l$, the transmit beamforming is determined by the ZF. According to constraint (\ref{zfmax1}a), $\boldsymbol w_k$ must lie in the orthogonal complement of the subspace $\mathrm{span}\{\boldsymbol h_{il}^\mathrm{H} \boldsymbol\Theta_l \boldsymbol G_l, \forall i\neq k,i\in\mathcal U_l\}$ and the orthogonal projector matrix on this  orthogonal complement is  
\begin{equation}\label{case1ZFeq1}
    \boldsymbol Z_k=\boldsymbol I -\boldsymbol G_k(\boldsymbol G_k^\mathrm{H}\boldsymbol G_k)^{\dagger}\boldsymbol G_k^\mathrm{H},
\end{equation}
where
\begin{equation}\label{case1ZFeq1_2}
    \boldsymbol G_k=\{\boldsymbol h_{il}^\mathrm{H} \mathrm{diag}(\boldsymbol \theta_{il}) \boldsymbol G_l \}_{\forall i\neq k,i\in\mathcal U_l}.
\end{equation}
As a result, the transmit beamforming for user $k$ is
\begin{equation}\label{case1ZFeq3}
    \boldsymbol w_k= \frac{\boldsymbol Z_k \boldsymbol h_{kl}^\mathrm{H} \boldsymbol\Theta_l  \boldsymbol G_l }{\left|\boldsymbol Z_k \boldsymbol h_{kl}^\mathrm{H} \boldsymbol\Theta_l  \boldsymbol G_l \right|}.
\end{equation}	
 

$\bullet$\quad\textbf{\emph{Frequency Division}}

The phase shift of the RIS can be given by \eqref{case1ZFeq0} since the RIS simultaneously serves all users in $\mathcal U_l$. The data rate of user $k$ is 
\begin{align}\label{case3zfmax1eq3}
    &\frac 1 {|\mathcal U_l|}B  \log_2\left(1+\frac{ |\mathcal U_l| p_k\left| 
    \boldsymbol h_{kl}^\mathrm{H} \boldsymbol\Theta_l  \boldsymbol G_l \boldsymbol w_k\right|^2}
    {\sigma^2}\right).
\end{align}
	
$\bullet$\quad\textbf{\emph{Time Division}}
	
{\color{black} The phase shift of the RIS can be given by \eqref{zfmax1eq2} since the RIS serves one user at each time.} The data rate of user $k$ is 
\begin{align}\label{case4zfmax1eq3}
    &\frac 1 {|\mathcal U_l|}B  \log_2\left(1+\frac{p_k\left| 
    \boldsymbol h_{kl}^\mathrm{H} \boldsymbol\Theta_l  \boldsymbol G_l \boldsymbol w_k\right|^2}
    {\sigma^2}\right).
\end{align}

\subsubsection{Multiple RISs Serve One User}
{\color{black} For the case that multiple RISs serve one user, the set of RISs is denoted by $\mathcal R_k$. Then, we can formulate the phase optimization problem as
\begin{subequations}\label{MRzfmax1}
    \begin{align}
        \max_{\boldsymbol\theta_l, \boldsymbol w_k} \quad&
        \left| \sum_{l\in\mathcal R_k }\boldsymbol h_{kl}^\mathrm{H} \mathrm{diag}(\boldsymbol \theta_l)   \boldsymbol G_l \boldsymbol w_k\right| \tag{\theequation}\\
        \textrm{s.t.} \quad
        &\theta_{ln} \in[0,2\pi], \quad \forall n\in\mathcal{N}_0,\\
        &\|\boldsymbol w_k\|=1,
    \end{align}
\end{subequations}

The transmit beamforming can be determined by
\begin{equation}
    \boldsymbol w_k=\frac{\sum_{l\in\mathcal R_k } \boldsymbol G_l^\mathrm{H}   \mathrm{diag}(\boldsymbol \theta_l)^\mathrm{H} \boldsymbol h_{kl}   }{\left|\sum_{l\in\mathcal R_k } \boldsymbol G_l^\mathrm{H}   \mathrm{diag}(\boldsymbol \theta_l)^\mathrm{H} \boldsymbol h_{kl}\right|}.
\end{equation}

For the phase shift, each $\boldsymbol \theta_l$ can be obtained by \eqref{zfmax1eq2}.}

\subsection{Power Control}
Given the RIS-user association, semantic compression ratio, and phase shift matrix in \eqref{sys2max0}, the power control problem can be formulated as
\begin{subequations}\label{sys2max0power}
    \begin{align}
        \max_{\boldsymbol p} \quad & \sum_{k=1}^K \frac{B}{\rho_k} \log_2\left(1+\frac{ p_k\left| 
        \sum_{l=1}^L x_{kl}\boldsymbol h_{kl}^\mathrm{H} \boldsymbol\Theta_l  \boldsymbol G_l \boldsymbol w_k\right|^2}
        {\sigma^2}\right),\tag{\theequation}\\
        \textrm{s.t.} \quad
        & \sum_{k=1}^K p_{k}\leq P, 
        \\& p_k\geq0,  \quad \forall k\in\mathcal{K},
    \end{align}
\end{subequations}
where the inter user interference is canceled since we adopt the phase shift design in the previous subsection. Since problem \eqref{sys2max0power} is convex, the optimal solution can be obtained by using the dual method.

\subsection{Overall Algorithm} 
The overall algorithm to solve problem \eqref{sys2max0} is given in Algorithm \ref{multipleuseropiAlo}. 

\begin{algorithm}[t]
    \caption{Joint Communication and Computation Optimization Algorithm for Problem \eqref{sys2max0}}
    \begin{algorithmic}[1]\label{multipleuseropiAlo}{\color{black}
        \STATE Initialize $\boldsymbol x^{(0)}, \boldsymbol p^{(0)}, \boldsymbol w^{(0)}, \boldsymbol \theta^{(0)}, \boldsymbol \rho^{(0)}$. Set iteration number $n=1$.
        \REPEAT
            \STATE With given $\boldsymbol p^{(n-1)}, \boldsymbol w^{(n-1)}, \boldsymbol \theta^{(n-1)}, \boldsymbol \rho^{(n-1)}$, obtain the RIS-user association $\boldsymbol x^{(n)}$ by using the many-to-many matching in Algorithm \ref{multipleusergreedy}.
            \STATE With given $\boldsymbol x^{(n)}, \boldsymbol p^{(n-1)}, \boldsymbol w^{(n-1)}, \boldsymbol \theta^{(n-1)}$, solve the semantic compression ratio subproblem using \eqref{scr} and obtain the solution $\boldsymbol \rho^{(n)}$.
            \STATE With given $\boldsymbol x^{(n)}, \boldsymbol p^{(n-1)}, \boldsymbol \rho^{(n)}$, solve the transmit and phase beamforming design subproblem and obtain the solution $\boldsymbol w^{(n)}, \boldsymbol \theta^{(n)}$.
            \STATE With given $\boldsymbol x^{(n)}, \boldsymbol w^{(n)}, \boldsymbol \theta^{(n)}, \boldsymbol \rho^{(n)}$, solve the power control subproblem and obtain the solution $\boldsymbol p^{(n)}$.
            \STATE Set $n=n+1$.
        \UNTIL{the objective value of problem \eqref{sys2max0} converges.}
        \STATE \textbf{Output}: The optimized $\boldsymbol x, \boldsymbol p, \boldsymbol w, \boldsymbol \theta, \boldsymbol \rho$.}
    \end{algorithmic}
\end{algorithm}

\section{Simulation Results}\label{section4}
\begin{figure}{\color{black}
    \centering
    \includegraphics[width=\linewidth]{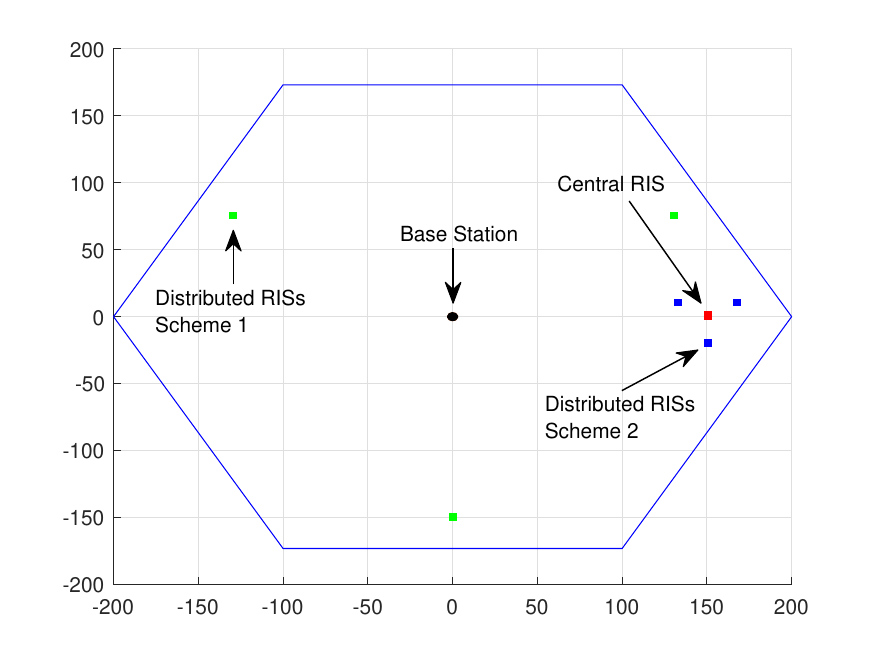}
    \caption{Three kinds of RIS distributions. Distributed RISs scheme 1 is colored in green, distributed RISs scheme 2 is colored in blue, and central RIS is colored in red.}
    \label{simfig01}}
\end{figure}

In this section, we provide the simulation results of the proposed algorithm. 
We consider a number of $K$ users uniformly distributed in a square area, as shown in Fig \ref{simfig01}.
The path loss model is $128.1+37.6\log_{10} d$ ($d$ is in km)
and the standard deviation of shadow fading is $8$ dB \cite{8352643}.
The bandwidth is $B=1$ MHz and the noise power spectral density is  $N_0=-174$ dBm/Hz.
Unless otherwise specified, we set
the number of elements in each RIS $N_0=121$.

\begin{figure}
    \centering
    \includegraphics[width=\linewidth]{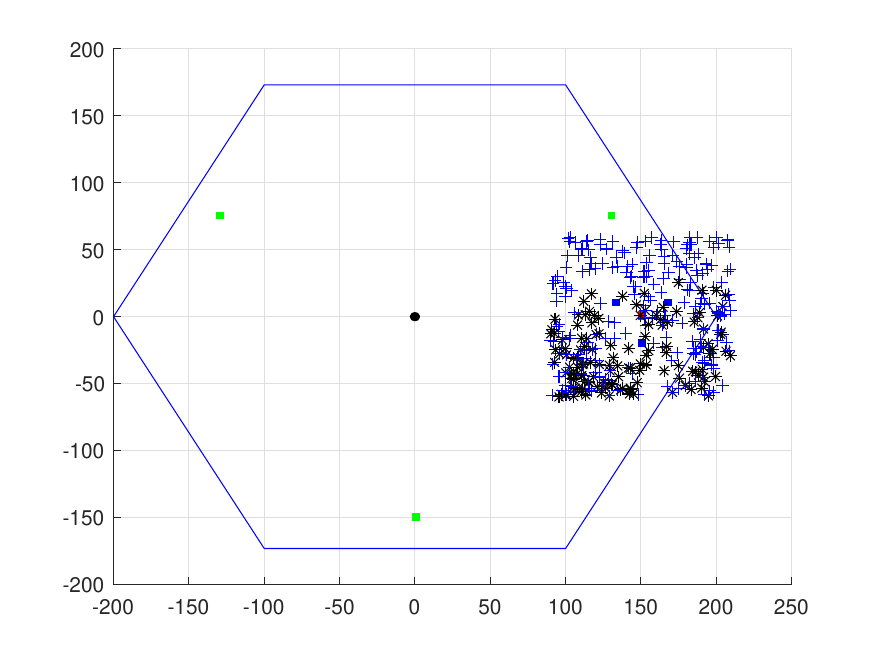}
    \caption{The user distribution case 1 around the central RIS.}
    \label{simfig02}
\end{figure}

\begin{figure}
    \centering
    \includegraphics[width=\linewidth]{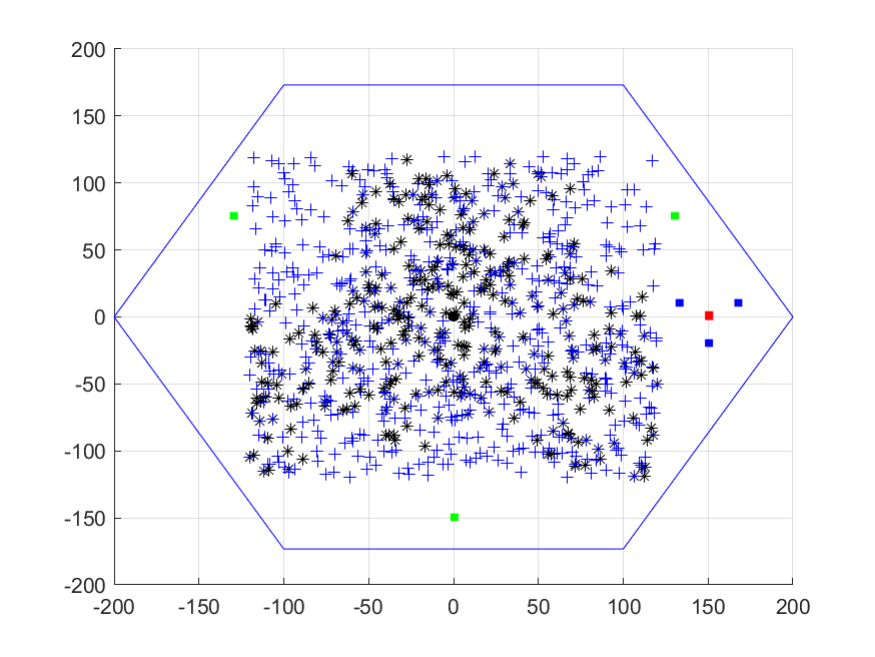}
    \caption{The user distribution case 2 in the whole space.}
    \label{simfig03}
\end{figure}

To show the effectiveness of distributed RISs, we consider two kinds of distributed RISs, distributed RISs scheme 1 with large distance between every two RISs, and distributed RISs scheme 2 with small distance between every two RISs. We also consider a central distribution with a large RIS, which has the same number of elements as two kinds of distributed RISs. 
Besides, Figs. \ref{simfig02} and \ref{simfig03} show two kinds of user distributions.

\begin{figure}
    \centering
    \includegraphics[width=\linewidth]{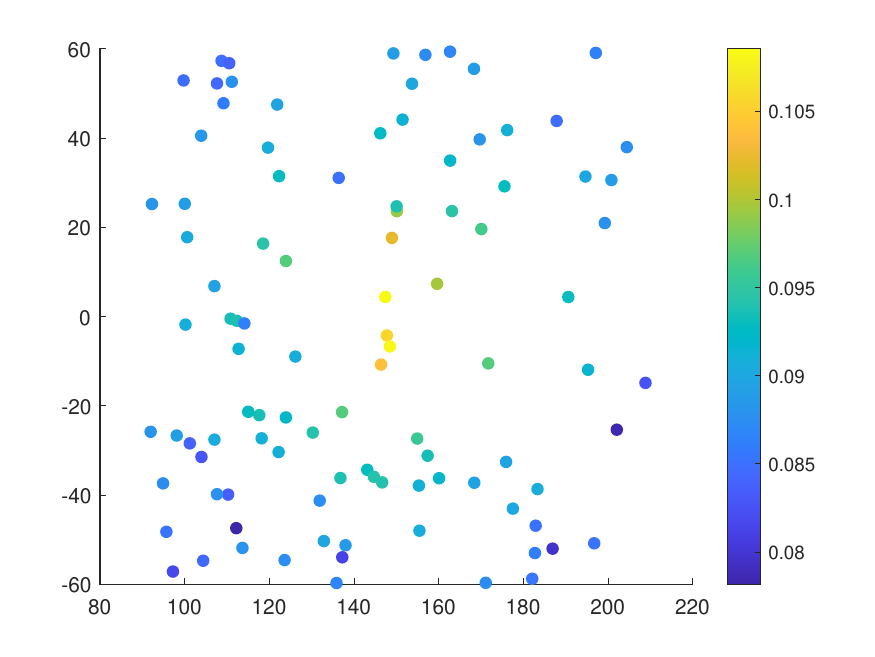}
    \caption{The rate distribution of central RIS under user distribution 1.}
    \label{simfig011}
\end{figure}

\begin{figure}
    \centering
    \includegraphics[width=\linewidth]{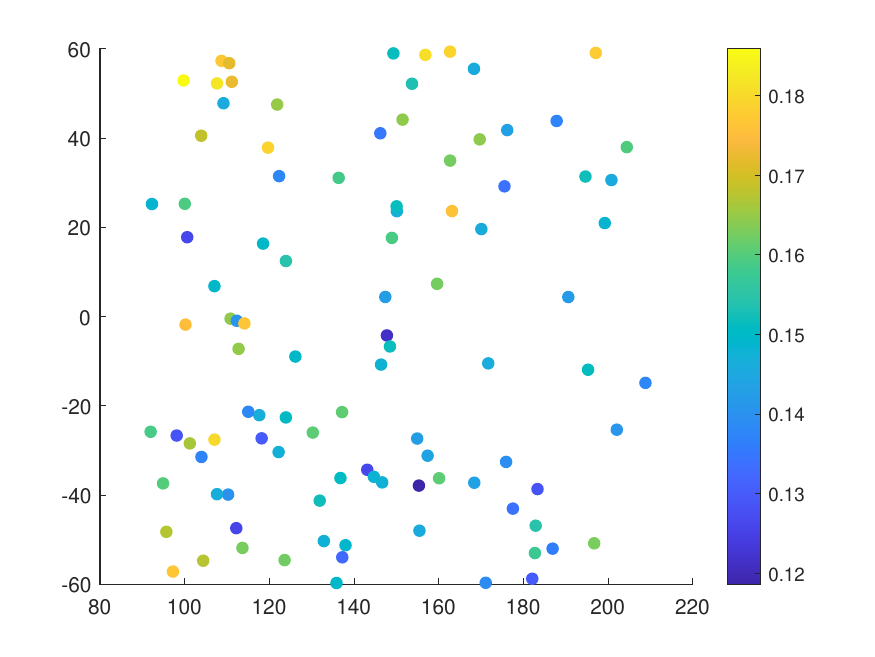}
    \caption{The rate distribution of distributed RISs scheme 1 under user distribution 1.}
    \label{simfig012}
\end{figure}

\begin{figure}
    \centering
    \includegraphics[width=\linewidth]{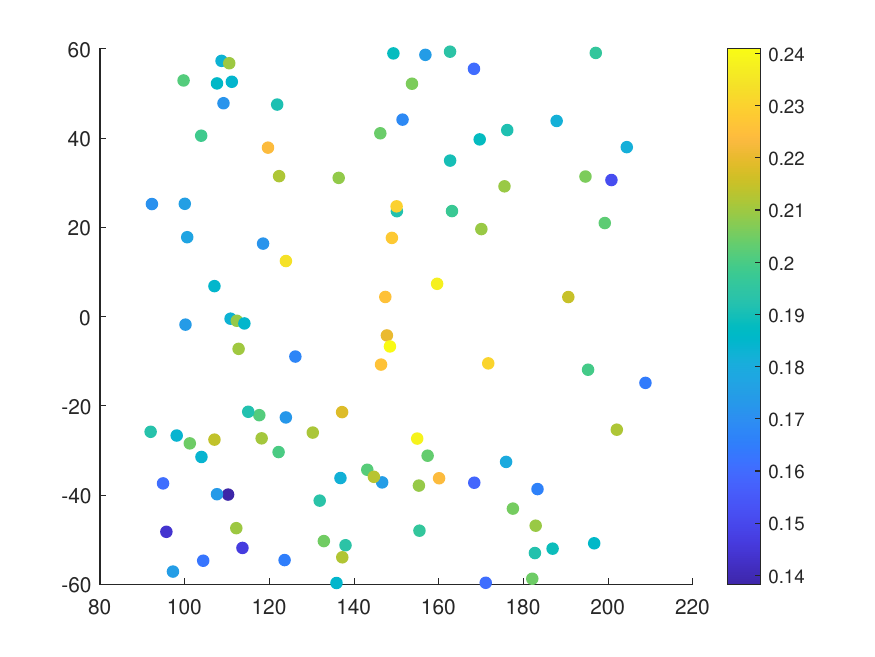}
    \caption{The rate distribution of distributed RISs scheme 2 under user distribution 1.}
    \label{simfig013}
\end{figure}

\begin{figure}
    \centering
    \includegraphics[width=\linewidth]{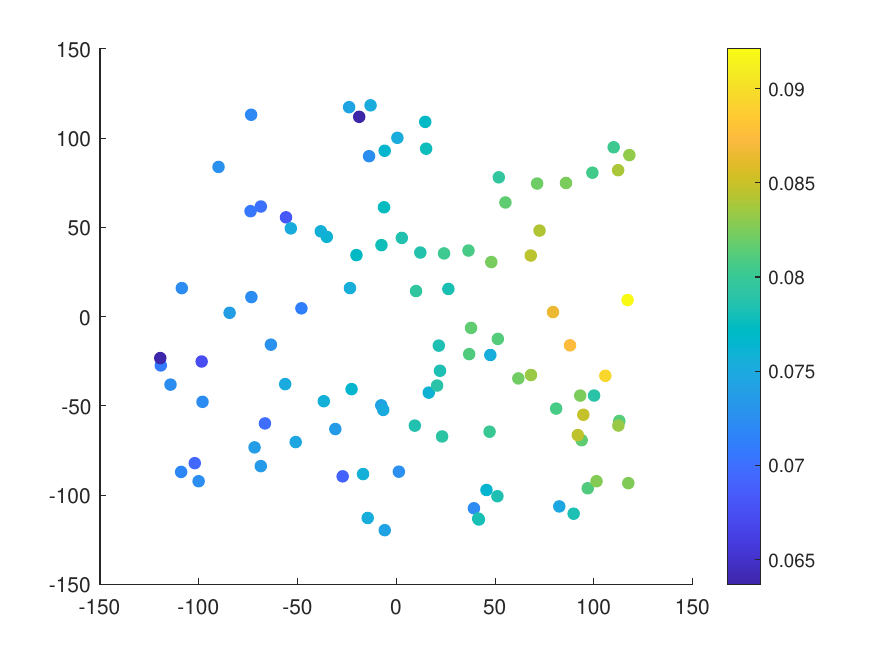}
    \caption{The rate distribution of central RIS under user distribution 2.}
    \label{simfig015}
\end{figure}

\begin{figure}
    \centering
    \includegraphics[width=\linewidth]{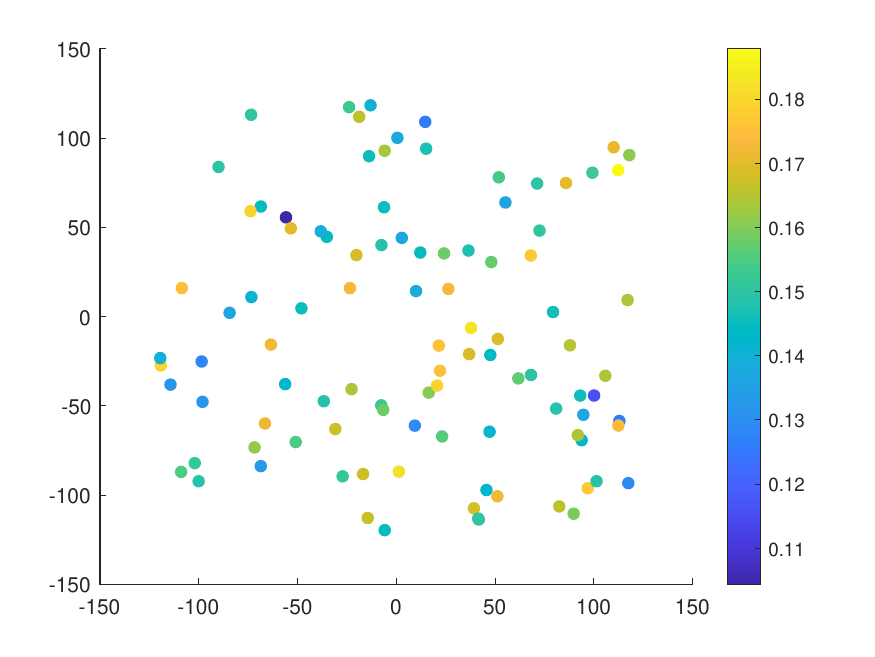}
    \caption{The rate distribution of distributed RISs scheme 1 under user distribution 2.}
    \label{simfig016}
\end{figure}

\begin{figure}
    \centering
    \includegraphics[width=\linewidth]{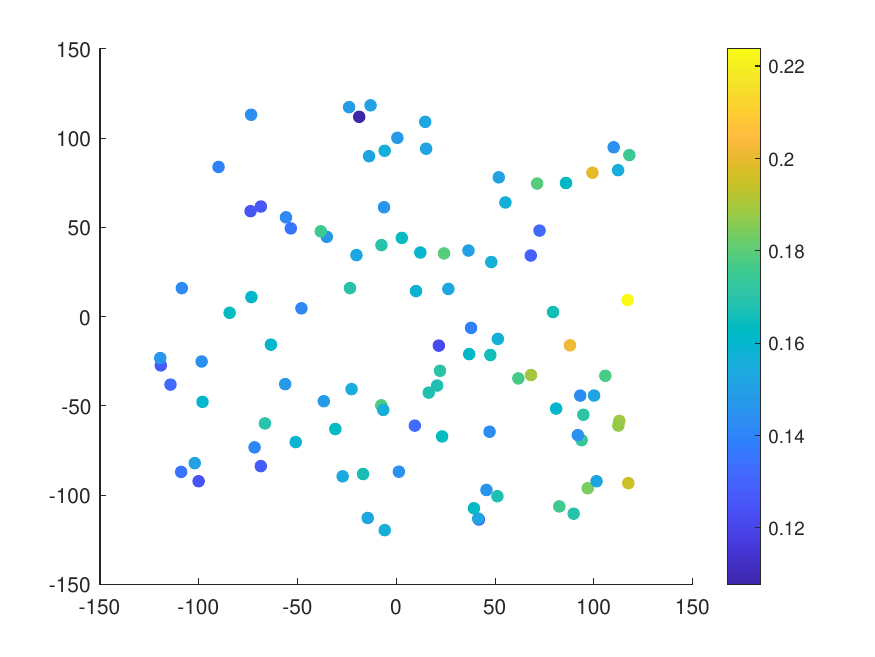}
    \caption{The rate distribution of distributed RISs scheme 2 under user distribution 2.}
    \label{simfig017}
\end{figure}

Figs.~\ref{simfig011}-\ref{simfig017} show the rate distributions of different RIS schemes under two kinds of user distributions. It can be found that the region of users with large rate is always higher of distributed RISs configurations compared to central RIS.

\begin{figure}{\color{black}
    \centering
    \includegraphics[width=\linewidth]{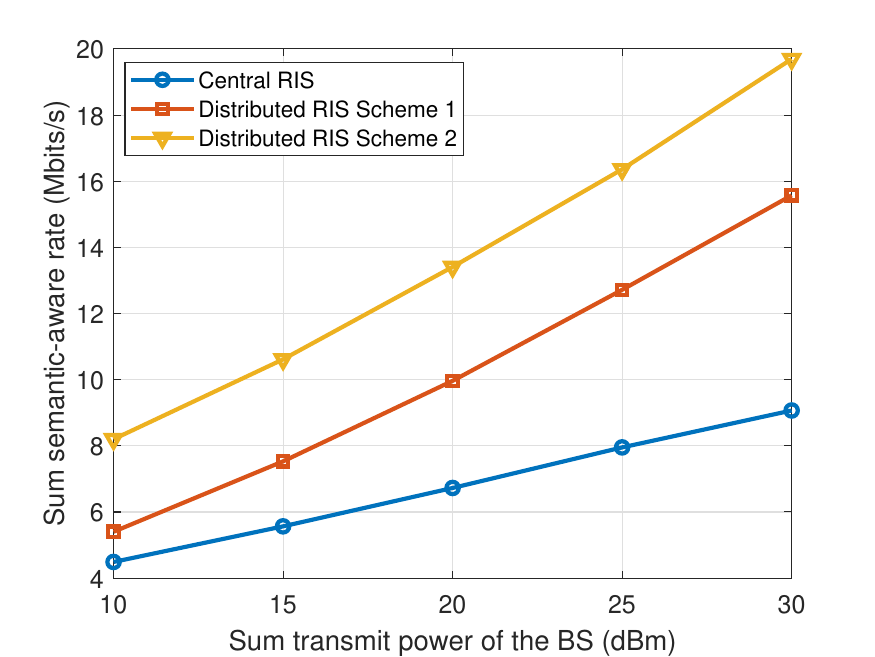}
    \caption{Sum semantic-aware rate versus sum transmit power under user distribution 1.}
    \label{simfig05}}
\end{figure}

\begin{figure}{\color{black}
    \centering
    \includegraphics[width=\linewidth]{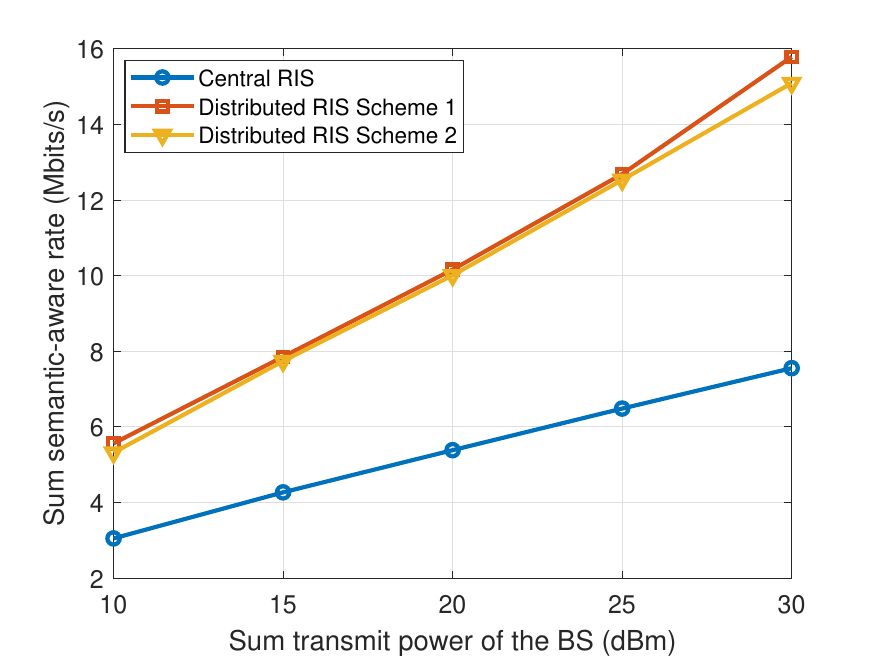}
    \caption{Sum semantic-aware rate versus sum transmit power under user distribution 2.}
    \label{simfig06}}
\end{figure}

Figs.~\ref{simfig05} and \ref{simfig06} show the sum rate versus sum transmit power under two different user distributions. It can be observed that both two distributed RISs schemes yield higher sum semantic-aware rate compared with the centralized RIS. For user distribution 1, it can be found that distributed RISs with small distance between every two RISs can achieve higher performance than the distributed RISs with large distance between every two RISs.
 
\begin{figure}{\color{black}
    \centering
    \includegraphics[width=\linewidth]{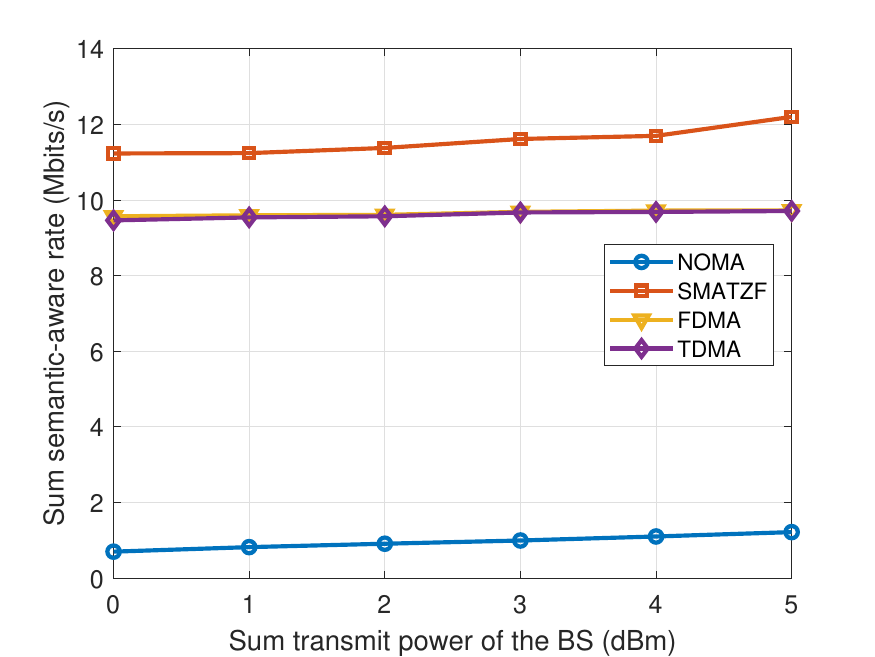}
    \caption{Sum semantic-aware rate versus transmit power of the BS.}
    \label{simfig1}}
\end{figure}

\begin{figure}{\color{black}
    \centering
    \includegraphics[width=\linewidth]{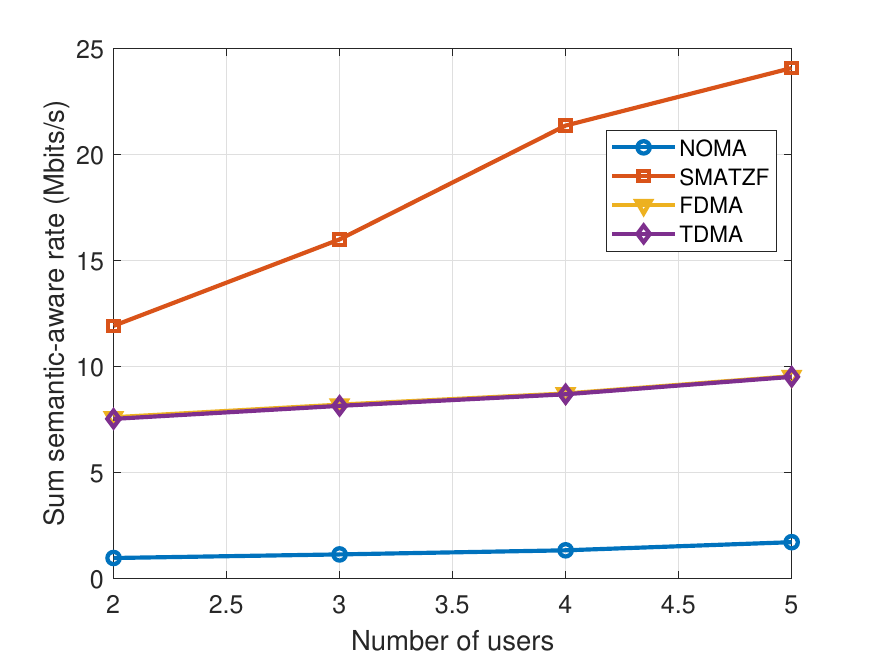}
    \caption{Sum semantic-aware rate versus number of users.}
    \label{simfig2}}
\end{figure}

\begin{figure}{\color{black}
    \centering
    \includegraphics[width=\linewidth]{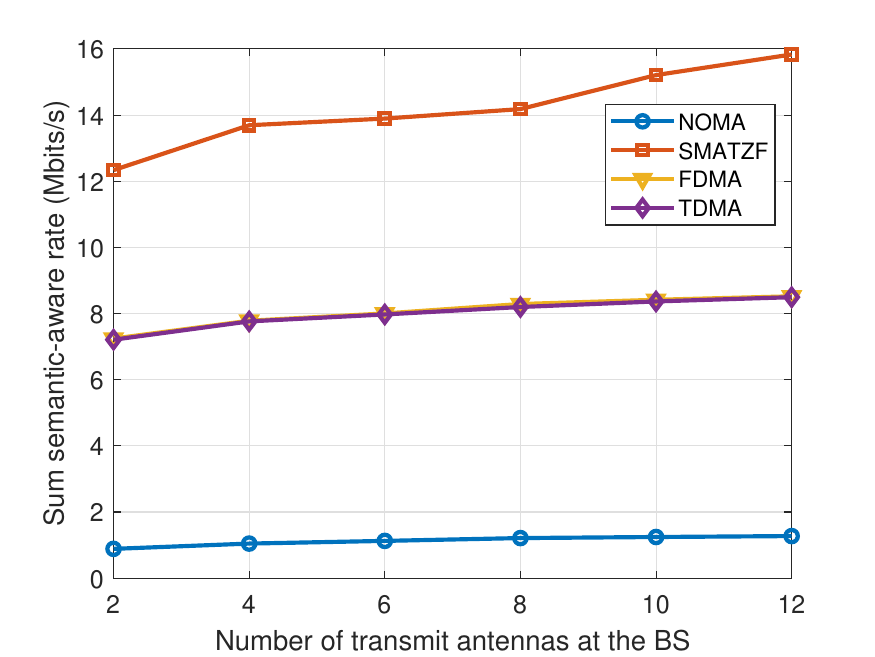}
    \caption{Sum semantic-aware rate versus number of antennas at the BS.}
    \label{simfig3}}
\end{figure}

In the following simulation, we set the number of users $K=10$, the total number of RISs $L=20$ in Figs.~\ref{simfig1}-\ref{simfig3}. 
According to Figs.~\ref{simfig1}-\ref{simfig3}, it is found that the space multiple access tensor ZF (SMATZF) achieves the best performance among all schemes. 

\section{Conclusions}\label{section5}
In this paper, we have investigated the spectral-efficient communication and computation resource allocation for RISs assisted PSC in IIoT. 
We formulated a joint communication and computation problem whose goal is to maximize the sum semantic-aware rate of the system under total transmit power, phase shift, RIS-user associate, and semantic compression ratio constraints. To solve this problem, a many-to-many matching scheme is proposed to solve the RIS-user association problem, the optimized semantic compression ratio is obtained with greedy policy, while the phase shift of RIS can be solved using the tensor based beamforming. 
Numerical results have illustrated the superiority of the proposed algorithm compared to the conventional schemes in terms of sum semantic-aware rate.

\bibliographystyle{IEEEtran}
\bibliography{main}

\end{document}